\documentclass[12pt,preprint]{aastex}

\def\amin{\ifmmode^{\prime}\else$^{\prime}$\fi}
\def\asec{\ifmmode^{\prime\prime}\else$^{\prime\prime}$\fi}

\begin{document}

\title{The UV, Optical, and IR Properties of SDSS Sources \\ Detected by {\it GALEX}}

\author{Marcel A. Ag\"ueros$^{1,2}$, \v{Z}eljko Ivezi\'c$^1$, Kevin R. Covey$^1$, Mirela Obri\'c$^3$, Lei Hao$^4$, Lucianne M. Walkowicz$^1$, Andrew A. West$^1$, Daniel E. Vanden Berk$^5$, Robert H. Lupton$^6$, Gillian R. Knapp$^6$, James E. Gunn$^6$, Gordon T. Richards$^6$, John Bochanski, Jr.$^1$, Alyson Brooks$^1$, Mark Claire$^1$, Daryl Haggard$^1$, Nathan Kaib$^1$, Amy Kimball$^1$, Stephanie M. Gogarten$^1$, Anil Seth$^1$, Michael Solontoi$^1$}

\altaffiltext{1}{Department of Astronomy, University of Washington, Box 351580, Seattle, WA 98195}
\altaffiltext{2}{NASA Harriett G. Jenkins Predoctoral Fellow; agueros@astro.washington.edu} 
\altaffiltext{3}{Kapteyn Institute, Postbus 800, Groningen 9700 AV, The Netherlands}
\altaffiltext{4}{Department of Astronomy, 610 Space Sciences Building, Cornell University, Ithaca, NY 14853}
\altaffiltext{5}{Department of Physics \& Astronomy, University of Pittsburgh, 3941 O'Hara Street, Pittsburgh, PA 15260}
\altaffiltext{6}{Princeton University Observatory, Princeton, NJ 08544}

\begin{abstract}
We discuss the ultraviolet, optical, and infrared properties of the Sloan 
Digital Sky Survey (SDSS) sources detected by the {\it Galaxy Evolution Explorer} 
({\it GALEX}) as part of its All--sky Imaging Survey (AIS) Early Release Observations (ERO). 
Virtually all ($>99$\%) of the {\it GALEX} sources in the overlap region are detected 
by SDSS; those without an SDSS counterpart within our $6\asec$ search radius are 
mostly unflagged {\it GALEX} artifacts. {\it GALEX} sources represent $\sim$$2.5\%$ of 
all SDSS sources within these fields and about half are optically 
unresolved. Most unresolved {\it GALEX}/SDSS sources are bright ($r < 18$) 
blue turn--off thick disk stars and are typically detected only in the 
{\it GALEX} near--UV band. The remaining unresolved sources include low--redshift 
quasars ($z < 2.2$), white dwarfs, and white dwarf/M dwarf pairs, and these 
dominate the optically unresolved sources detected in both {\it GALEX} bands. 

Almost all the resolved SDSS sources detected by {\it GALEX} are fainter than the 
SDSS ``main'' spectroscopic limit (conversely, of the SDSS 
galaxies in the ``main'' spectroscopic sample, about $40$\% are detected in 
at least one {\it GALEX} band). These sources have colors consistent with those of 
blue (spiral) galaxies ($u - r < 2.2$), and most are detected in both {\it GALEX} 
bands. Measurements of their UV colors allow much more accurate and robust 
estimates of star--formation history than are possible using only SDSS data. 
Indeed, galaxies with the {\it most recent} ($\la20$ Myr) star formation can be robustly 
selected from the {\it GALEX} data by requiring that they be brighter in the 
far--ultraviolet than in the near--ultraviolet band. However, older starburst 
galaxies have UV colors similar to AGN, and thus cannot be selected 
unambiguously on the basis of {\it GALEX} fluxes alone. Additional information, 
such as spatially resolved far--UV emission, optical morphology, or X--ray and 
radio data, is needed before the blue {\it GALEX} color can be unambiguously 
interpreted as a sign of recent star formation.
 
With the aid of Two Micron All Sky Survey (2MASS) data, we construct and 
discuss median 10 band UV--optical--infrared spectral energy distributions 
for turn--off stars, hot white dwarfs, low--redshift quasars, and spiral and 
elliptical galaxies. We point out the high degree of correlation between the UV 
color and the contribution of the UV flux to the UV--optical--infrared flux of 
galaxies detected by {\it GALEX}; for example, this correlation can be used to predict 
the SDSS $z$ band measurement, using only two {\it GALEX} fluxes, with a scatter of 
only 0.7 magnitudes. 
\end{abstract}

\keywords{catalogs ---  galaxies: active ---  galaxies: starburst --- ultraviolet: galaxies --- ultraviolet: stars --- ultraviolet: general}

\section{                         Introduction                    }
\begin{figure} 
\includegraphics[bb=13 13 230 140, width=0.75\columnwidth]{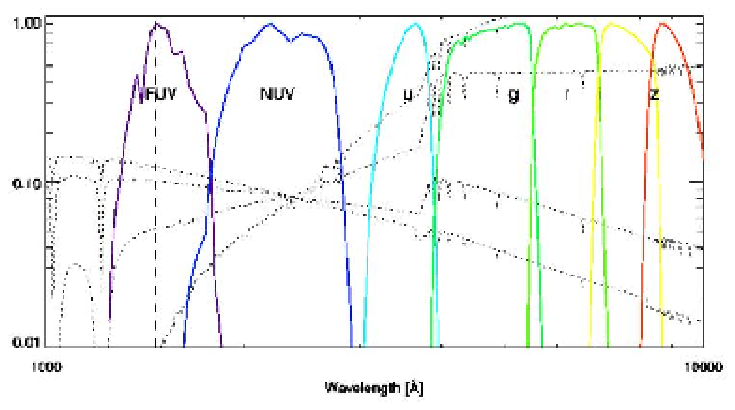}
\caption{The {\it GALEX} FUV and NUV and the SDSS {\it u, g, r, i, z} filters. The dashed lines correspond to the spectra of galaxies with different starburst histories. From ${\rm http://www.galex.caltech.edu/EROWebSite/Early\_release\_data\_description\_part2.htm}$.}
\label{bands} \end{figure}

Launched in April 2003, the {\it Galaxy Evolution Explorer} ({\it GALEX}) made its 
first public data release (the Early Release Observations, or ERO) at the end
of 2003. Included in the ERO are fields from several different {\it GALEX} surveys 
that overlap with the Sloan Digital Sky Survey (SDSS; York et al.~2000), 
allowing one to study sources over 
nearly the entire $1000$ to $10000\ {\rm \AA}$ range (see Fig.~\ref{bands}). 
Here we report the results of matching {\it GALEX} All--sky Imaging Survey 
(AIS; $t_{exp} \approx 100$~s) observations with SDSS data in the overlapping 
fields. There are other, deeper {\it GALEX} observations of SDSS fields in the ERO, 
but AIS is the {\it GALEX} survey that will eventually provide the largest sky 
coverage\footnote{The Medium Imaging Survey is $\sim$$2.5$ magnitudes deeper, and will cover about 1000 deg$^2$ of sky, overlapping the SDSS footprint.}. 
It is therefore the most appropriate {\it GALEX} survey for discussing the bulk 
properties of objects in the overlapping {\it GALEX}/SDSS region. 

Here we describe the optical properties of matched {\it GALEX}/SDSS sources in 
three AIS ERO fields, covering $\sim$$3$ deg$^2$ of sky, which overlap
the SDSS DR1 footprint. The first full {\it GALEX} public data release should 
contain about $1000$ deg$^2$ of overlap with the SDSS (Seibert et al.~2005a) 
and therefore allow the construction of a much larger sample of matched 
objects than discussed in this paper. However, even 
the fairly small sample discussed here (about 3000 matched sources) is 
sufficient to highlight some of the challenges in producing 
a good sample of {\it GALEX}/SDSS sources, and to characterize the optical SDSS 
properties of the matched sources---as well as to produce representative 
spectral energy distributions (SEDs) for stars, quasars, and galaxies 
detected by these two surveys and by the Two Micron All Sky Survey (2MASS). 

In the next section we briefly describe the three surveys we used in this work. 
Section~3 describes the process of matching {\it GALEX}/SDSS objects and of producing
a clean photometric sample of matched objects. It also includes a discussion of 
{\it GALEX} objects without SDSS counterparts, as well as an analysis of the 
repeatability of {\it GALEX} measurements. Section~4 presents an analysis of the 
optical properties of unresolved and resolved {\it GALEX}/SDSS sources, a discussion 
of the SEDs of a number of interesting classes of sources, and an estimate of 
the UV contribution to the UV--optical--infrared flux of galaxies. We discuss the 
significance of our results in Section~5, and in particular compare them to 
those in the recently published Yi et al.~(2005) and Rich et al.~(2005) studies
of star formation in early--type galaxies detected by {\it GALEX}.

\section{                    Observations                            }

{\it GALEX} will eventually map the entire sky at wavelengths between $1344$ and 
$2831$ ${\rm \AA}$ in two bands: the near ultraviolet (NUV; 
$\lambda_{eff} = 2271 {\rm \AA}, \lambda / \Delta\lambda = 90$) and the far 
ultraviolet (FUV; $\lambda_{eff} = 1528 {\rm \AA}, \lambda / \Delta\lambda = 200$). 
When comparing positions to the Tycho--2 catalog (H$\o$g et al.~2000), 
$80\%$ of {\it GALEX}--detected stars are found within $1.5\asec$ in the NUV and 
$2.8\asec$ in the FUV of their expected positions (Morrissey et al.~2005). 
{\it GALEX}'s $0.5$ m telescope and $1.2^{\rm o}$ field of view will also be used to 
make deep observations
($>$ tens of kiloseconds) of individual fields of interest, such as the
Lockman Hole and the {\it Chandra} Deep Field--South. The mission's primary
science goals are to observe star--forming galaxies and to track galaxy 
evolution (Martin et al.~2005). The {\it GALEX} Early Release Observations (ERO) 
include 10 fields, three of which are AIS observations that overlap the 
Sloan Digital Sky Survey (SDSS) footprint. The AIS fields were observed for 
113, 111, and 113 seconds respectively, and each covers $1.2$~deg$^2$ (the fields
overlap slightly, however, so that the total area on the sky is smaller; see 
Fig.~\ref{onsky}). 
While for most classes of objects in SDSS the SEDs drop off quickly in the UV, 
the ERO fields are observed to $n \sim 22$ (NUV) and $f \sim 22$ (FUV)\footnote{We 
use $f$ and $n$ to denote {\it GALEX} AB magnitudes in the far-- and near--ultraviolet 
bands, respectively.}, 
deep enough that we expect to find {\it GALEX} counterparts for a large number of 
SDSS sources. 

The Sloan Digital Sky Survey is currently mapping one quarter of the sky 
at optical wavelengths. SDSS uses a dedicated 2.5 m telescope at the Apache 
Point Observatory, New Mexico, to produce homogeneous five color 
{\it u, g, r, i, z} CCD images to a depth of $r\sim22.5$ (Fukugita et al.~1996;
Gunn et al.~1998; Smith et al.~2002; Hogg et al.~2002) accurate to $0.02$ 
magnitudes (both absolute calibration, and root--mean--square scatter for 
sources not limited by photon statistics; Ivezi\'{c} et al.~2004). 
Astrometric positions are accurate to better than $0.1\asec$ per coordinate
(rms) for sources with $r<20.5$ (Pier et al.~2003), and the morphological
information from the images allows reliable star/galaxy separation to 
$r\sim 21.5$ (Lupton et al.~2002). The survey's coverage of 
$\sim$$10^4$~deg$^2$ in the North Galactic Cap and of $\sim$$200$~deg$^2$ 
in the Southern Galactic Hemisphere will result in photometric measurements 
for over $10^8$ stars and a similar number of galaxies.
Additionally, SDSS will obtain spectra for over $10^6$ objects, 
including $10^6$ galaxies and $10^5$ quasars. The third public Data 
Release (DR3) includes imaging data for $5282$ deg$^2$ of sky, and catalogs 
$1.4 \times 10^8$ objects (Abazajian et al.~2005).

Finally, in constructing UV--optical--infrared SEDs for our UV--selected sample
of objects, we also utilize data from the Two Micron All Sky Survey (2MASS) survey. 
2MASS used two 1.3 m telescopes to survey the entire sky in near--infrared 
light\footnote{See http://www.ipac.caltech.edu/2mass.}. 
Each telescope's camera was equipped with three $256\times256$ arrays of 
HgCdTe detectors with $2\asec$ pixels and observed simultaneously in 
the $J$ (1.25 $\mu{\rm m}$), $H$ (1.65 $\mu{\rm m}$), and $K_s$ 
(2.17 $\mu{\rm m}$) bands.
The detectors were sensitive to point sources brighter than about 1 mJy at 
the $10\sigma$ level, corresponding to limiting (Vega--based) magnitudes of 
$15.8$, $15.1$, and $14.3$, respectively. Point source photometry is repeatable 
to better than $10\%$ precision at this level, and the astrometric 
uncertainty for these sources is less than $0.2\asec$. The 2MASS
catalogs contain positional and photometric information for $\sim$$5\times10^8$ 
point sources and $\sim$$2\times10^6$ extended sources. Finlator et al.~(2000)
and Ivezi\'{c} et al.~(2001) describe the properties of sources detected by 
both SDSS and 2MASS (in particular, Fig.~3 in Finlator et al.~compares the
SDSS and 2MASS bandpasses, and is analogous to Fig.~\ref{bands} in this paper).

\section{                     Matching {\it GALEX} and SDSS                     }
\subsection{Positional Offsets}
\begin{figure} 
\includegraphics[bb=15 15 440 560, width=0.75\columnwidth, angle=90]{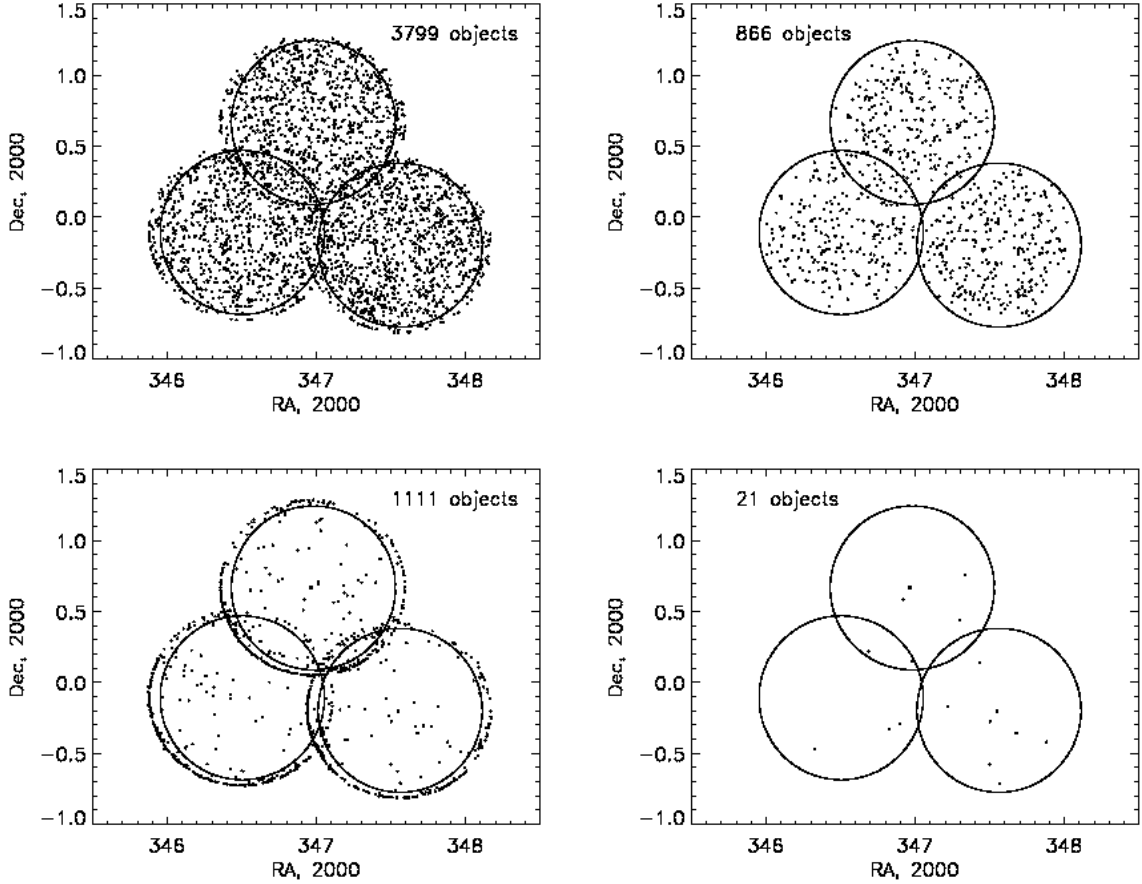}
\caption{The distribution of both matched and unmatched {\it GALEX} sources; matches imply an SDSS source cataloged within 6$\asec$ of the {\it GALEX} position. The top left panel shows all {\it GALEX} sources with an SDSS counterpart; the bottom left panel is {\it GALEX} sources without an SDSS counterpart. The right panels are obtained when several quality cuts are applied to the {\it GALEX} and SDSS data; the top right panel is for matched sources, and the bottom right panel for unmatched sources. The circles in all four panels represent the R~$= 0.55^{\rm o}$ field of view for which {\it GALEX} astrometry is most accurate.} 
\label{onsky} \end{figure}

The astrometry for both the {\it GALEX} and SDSS surveys is sufficiently accurate 
that the typical astrometric errors are much smaller than the average source 
separation; this significantly simplifies the matching algorithm. 
We began by correlating the {\it GALEX} source positions with positions  
in the SDSS catalog, taking a $6\asec$ matching radius. This 
corresponds to the full--width--at--half--maximum angular
resolution for the NUV channel (Morrissey et al.~2005). Fig.~\ref{onsky} 
illustrates the results of this matching: of the $4910$ UV--detected 
objects in the three {\it GALEX} fields, we find optical counterparts for $3799$ 
($77\%$) sources, of which $686$ ($18\%$) are saturated in the optical.
About 5\% of matched {\it GALEX} sources have more than one SDSS counterpart\footnote{Note that
this fraction of multiply matched {\it GALEX} sources is somewhat lower than that reported by Seibert et al.~(2005a).}. This is consistent with random matching,
based on the mean separation between two SDSS sources of $\sim$$30\asec$. 
In these cases, we simply take the closest match for evaluating sample 
completeness, and limit the matching radius to $3\asec$ when studying 
colors of matched sources in \S 4 (for this matching radius, less than 1\% of
{\it GALEX} matches have more than one SDSS counterpart).

A closer look at the matches---and especially the non--matches, those {\it GALEX} 
objects without an SDSS counterpart---shows clear structure in the pattern of 
matching (see the lower left panel in Fig.~\ref{onsky}). Objects
along the edges of the {\it GALEX} field of view are far more likely not to have 
an optical counterpart. This is due mainly to distortions in the {\it GALEX} 
fields and to problems in the flat--fielding along the field edges; as a 
result, many spurious sources are detected by the {\it GALEX} data analysis 
pipeline (T.~Wyder 2004, private communication). To avoid 
this contamination, we select an inclusion distance from the {\it GALEX} field
center of R~$\leq0.55^{\rm o}$, which defines the size 
of the effective area of each of the three fields overlapping with SDSS. 
We then have $3007$ {\it GALEX} sources with SDSS counterparts, and only $192$ 
without a match within $6\asec$, or $94\%$ and $6\%$, respectively, of the 
total number of {\it GALEX} sources within the area defined above.

\begin{figure} \plotone{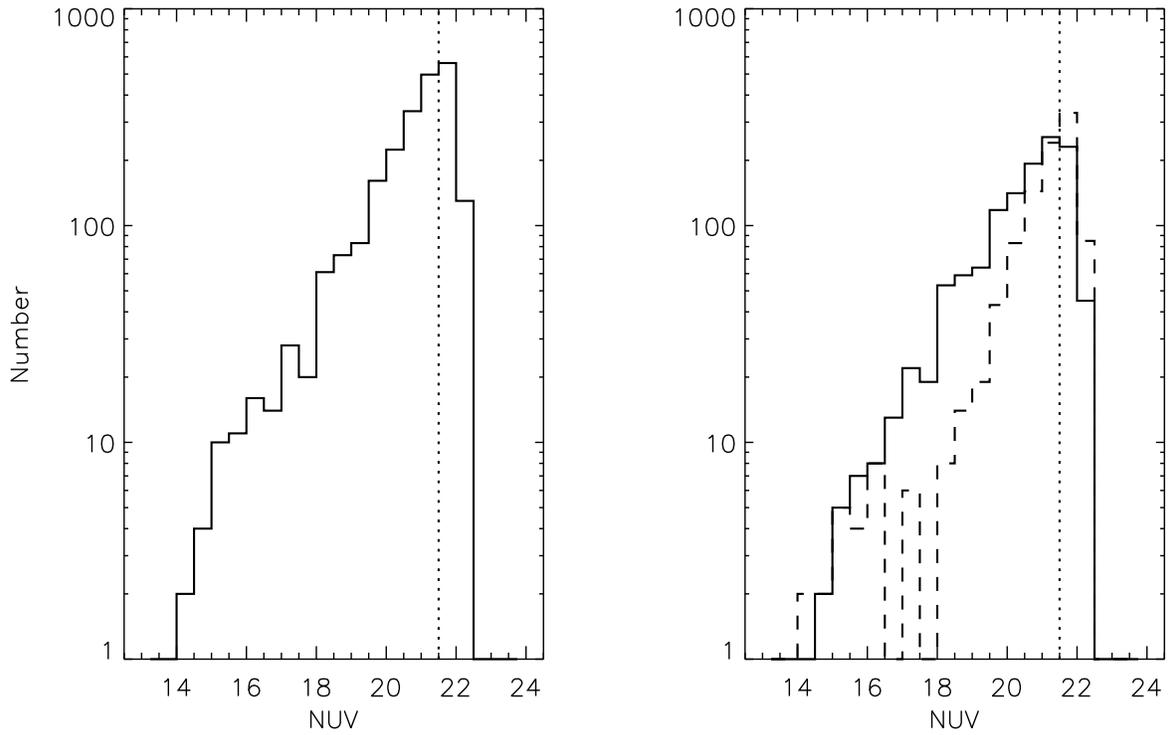}
\caption{The left panel shows the differential $n$ magnitude distribution for the unflagged {\it GALEX} sources with SDSS counterparts. We adopt $n = 21.5$ magnitudes as the {\it GALEX} faint completeness limit. The right panel shows the distribution separately for the optically unresolved (solid) and resolved (dashed) sources.}
\label{limits} \end{figure}

Further cuts are then applied to the data to obtain the highest
quality sample of {\it GALEX}/SDSS sources. We determined the {\it GALEX} faint 
completeness limit from a histogram 
of the $n$ magnitudes of {\it GALEX} sources with an SDSS counterpart and within 
$0.55^{\rm o}$ of their respective field centers (Fig.~\ref{limits}).
{\it GALEX} sources begin to drop out at $n$ $\gtrsim22$ magnitudes; we
select $n$ $=21.5$ as a conservative completeness limit for our sample. 
For the optical counterparts, we require $14< g <22$. Furthermore, we apply 
a number of conditions based on
data processing flags in the two data sets. We require that the optical 
counterpart be a unique detection and not saturated in SDSS (for details 
see Stoughton et al.~2002). We also require that the {\it GALEX} artifact flag be 
set to zero for both the near--UV and far--UV detections. Bright star halos 
appear to be one of the major sources of artifacts in both the NUV and FUV 
{\it GALEX} data sets, while other problems (dichroic ghosts or detector hot spots, 
for example) tend to affect preferentially one set of detections or the 
other\footnote{See ${\rm http://www.galex.caltech.edu/EROWebSite/Early\_release\_data\_description\_part3.htm}$ 
for a full description of {\it GALEX} image artifacts.}.
 
\begin{figure} \plotone{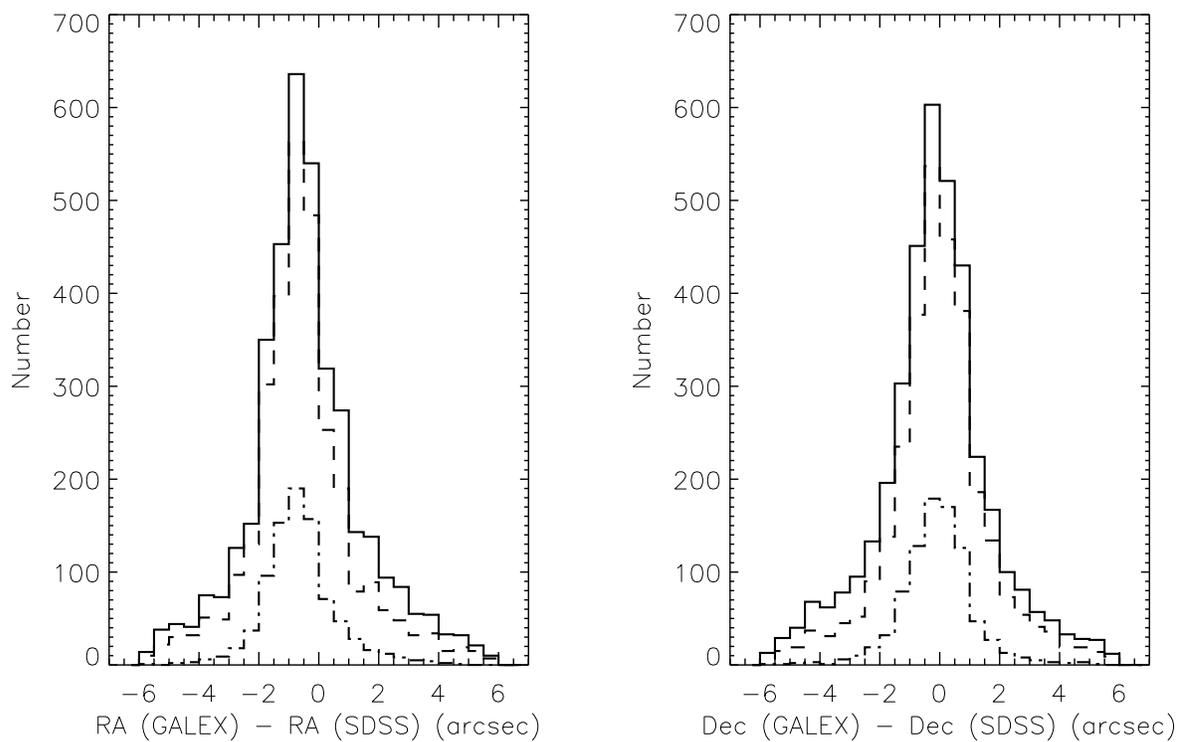} 
\caption{The distribution of positional offsets for {\it GALEX} sources with SDSS counterparts. The solid line is for all 3799 matches. The dashed line is for the 3007 matched objects less than $0.55^{\rm o}$ from the center of the {\it GALEX} field, while the dot--dashed line is for the 866 objects satisfying a number of photometric criteria in both surveys and constituting our cleanest sample of matches. The median values and the root--mean--square scatter for these distributions are listed in Table 1.} 
\label{offsets} \end{figure}

The sky distribution of the resulting sample of $866$ matched, ``clean'' 
sources is shown in Fig.~\ref{onsky} (top right panel). Table 1 gives the 
median astrometric offsets and standard deviations for each of the three 
{\it GALEX} fields, and for the overall list of matched sources; for comparison, 
the offsets obtained during all three of the matching procedures described 
above are included (i.e., for all matched {\it GALEX}/SDSS
sources, for matches with R~$\leq 0.55^{\rm o}$, and for clean matches).
Fig.~\ref{offsets} illustrates these results.

We note that eliminating matches based on the {\it GALEX} NUV/FUV flags 
strongly impacts the spatial distribution of acceptable matches, so 
that there now seems to be a dearth of clean sources near the edges of the 
R~$\leq 0.55^{\rm o}$ disks. This suggests that perhaps the {\it GALEX} flags are
in fact too conservative, and that we are losing good matches in these regions. 

\subsection{Unmatched {\it GALEX} Sources}
\begin{figure} 
\includegraphics[bb=16 15 750 390, width=0.99\columnwidth]{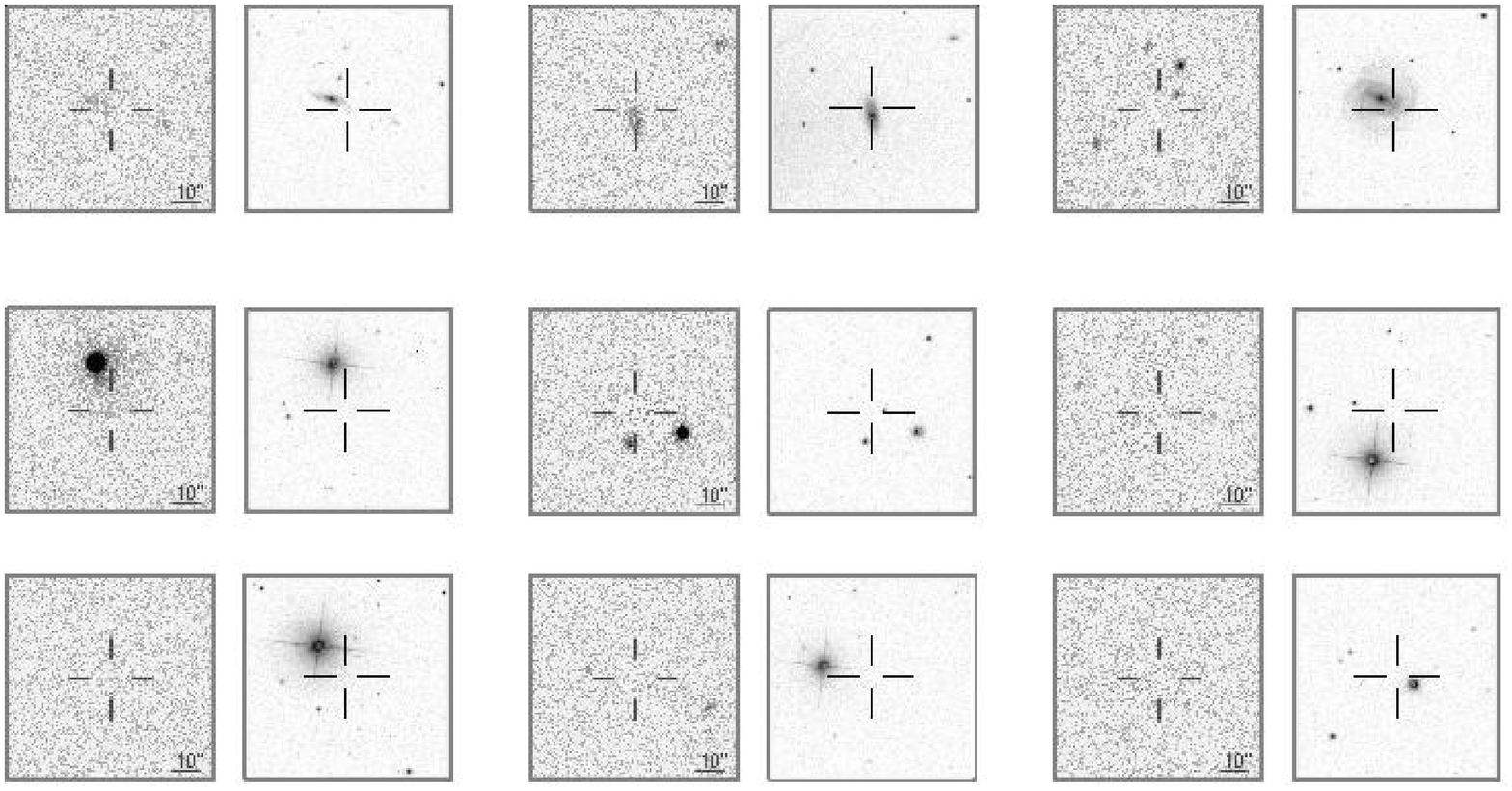}
\caption{NUV band and composite {\it g, r, i} SDSS images of the ``explained'' unmatched {\it GALEX} sources (the first nine sources listed in Table 2, in the same order, from top left to bottom right). For each source, the {\it GALEX} NUV band image is to the left and the SDSS image is to the right; the images are $150\asec$ on a side, with equivalent resolution (the {\it GALEX} scale bar is $20\asec$, not $10\asec$; S. Salim 2005, private communication). In all the images the cross hairs indicate the quoted position of the {\it GALEX} source, and North is up and East to the left. The three sources in the top row are most likely associated with the galaxies shown in the optical images. The six sources in the bottom two rows are likely to be missed artifacts---false detections due to nearby bright star halos.
\label{no_1}}
\end{figure}

\begin{figure} 
\vspace{-1in}
\includegraphics[bb=16 15 745 650, width=0.99\columnwidth]{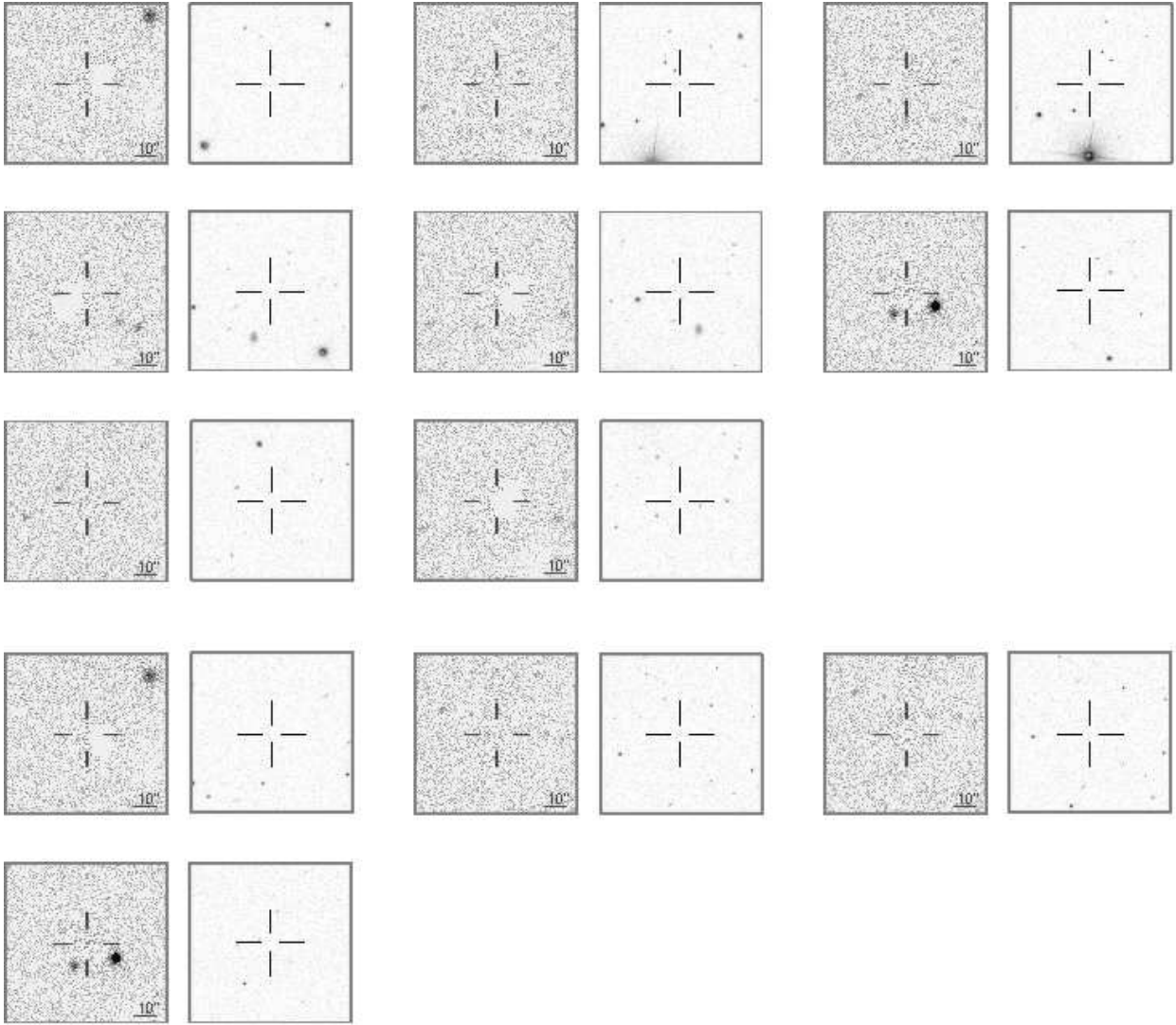}
\caption{NUV band and composite {\it g, r, i} SDSS images of the {\it GALEX} sources (the last 12 sources in Table 2) without SDSS counterparts that are more difficult to interpret. The top three rows show possible stellar artifacts; here however the bright star is more than $1\amin$ from the quoted {\it GALEX} position. As for the four sources whose {\it GALEX} and SDSS images are shown in the bottom two rows, their nature remains unknown. Only the last one, J231131.21$-$002510.96, was flagged as suspect by the extraction pipeline.
\label{no_2}} 
\end{figure}

Interestingly, it appears that a handful (21) of {\it GALEX} sources 
have no SDSS counterparts within $6\asec$, even when highly restrictive quality
cuts are applied. These sources are listed in Table 2, and their positions are 
plotted in the bottom right panel in Fig.~\ref{onsky}. We used the 
Multimission Archive at the Space Telescope Science Institute (MAST\footnote{${\rm http://galex.stsci.edu.}$ MAST is operated by AURA under grant NAG5--7584.}) and the SDSS Image List Tool\footnote{${\rm http://cas.sdss.org/astro/en/tools/chart/list.asp}$.} to examine the {\it GALEX} and SDSS ``postage stamp'' images for 
all 21 sources (see Fig.~\ref{no_1} and Fig.~\ref{no_2} for mosaics of these 
images). 

\begin{itemize}
\item {\bf Extended galaxies}

The only two {\it GALEX} sources without an SDSS counterpart detected both in the NUV
and FUV, J230734.52$-$001731.04 and J230919.65$+$004515.64, are associated with 
optically large SDSS galaxies whose centers are farther than $6\asec$ from the 
{\it GALEX} position---this large separation explains why they were not matched (for 
SDSS J230920.2$+$004523.3, the counterpart to J230919.65$+$004515.64, a 
spectrum is available; this is clearly an emission--line galaxy). In addition, 
J230644.65$+$001302.13, detected only in the NUV band, also appears to be 
associated with a galaxy, although here the {\it GALEX} source is positioned on the 
very edge of the optically detected galaxy (see the top right pair of images in 
Fig.~\ref{no_1}).

In all of these cases the {\it GALEX} source extraction pipeline (based on 
SEXtractor; Bertin \& Arnouts 1996) did not label the sources as
artifacts, but did set the extraction flag to 3 in all the bands in which it 
claimed a detection, indicating that the object was originally blended. 
These detections are supported by the recent work of Thilker et al.~(2005), 
who observe significant {\it GALEX} emission at large radii in nearby galaxies. 
Our off--center detections may be UV emission coming from star--forming regions 
at large galactic radii, similar to those found in the tidal tails of 
``the Antennae'' merging system (Hibbard et al.~2005).

\item {\bf Artifacts}

Several other sources appear to be close enough to bright stars that they may 
in fact be artifacts that were not flagged. J230518.70$-$002816.29, 
J230751.11$+$003936.81, and J230959.96$-$003441.17 were all flagged by 
SEXtractor as either having bright neighbors close enough to bias the 
photometry (flag $= 1$), or as having originally been blended sources (flag $= 2$). 
While J230717.62$-$001853.40, J230852.36$-$001005.47, and J231042.50$-$002126.92 
were not flagged at all, their SDSS images suggest that they could indeed be detections 
due to bright star halos (see the bottom two rows of Fig.~\ref{no_1}).
 
We note that J230519.28$-$002741.34 (GSC 05242$-$00801; $m_B = 11.4$, 
$m_V = 11.0$\footnote{This research has made use of the SIMBAD database, 
operated at CDS, Strasbourg, France.}), the star responsible 
for the halo detected as J230518.70$-$002816.29, is very bright in the 
near--UV: $n = 15.09\pm0.01$ ($f = 20.38\pm0.21$). 

An additional eight {\it GALEX} sources are found between $1\amin$ and $3\amin$ from 
SDSS--detected stars with $r < 13.5$ (see the top three rows of Fig.~\ref{no_2};
three of these sources were flagged by SEXtractor as having originally been 
blended). 

There are 1537 {\it GALEX} sources detected less than $0.55^{\rm o}$ from their 
respective field centers with $n < 21.5$ and no $n$ or $f$ flags. Assuming 
that the 14 sources described here are stellar artifacts, we can place an 
upper limit of $1\%$ for the fraction of the ``photometric'' {\it GALEX} sources 
that are unflagged artifacts.  

\item {\bf Unexplained non--matches}

Four {\it GALEX} sources do not have a bright star within a few arcminutes (see the 
bottom two rows of Fig.~\ref{no_2}). J231131.21$-$002510.96 is the only one 
flagged by SEXtractor as having been deblended, suggesting that it is an 
artifact. However, nothing in the MAST provides any explanation for the nature 
of the other non--matches. These mysterious sources represent fewer than 
$0.3\%$ of the total number of photometric {\it GALEX} sources within 
$0.55^{\rm o}$ of the field centers. They do not have counterparts within 
$30\asec$ cataloged in either SIMBAD or NED\footnote{This research has made use of the NASA/IPAC Extragalactic Database (NED) which is operated by the Jet Propulsion Laboratory, California Institute of Technology, under contract with the National Aeronautics and Space Administration.}, 
suggesting that they may not be real sources. On the other hand, if their 
UV detections could be confirmed, they would represent an interesting class 
of extremely blue (UV--to--optical) sources. A larger sample of {\it GALEX} sources 
may indeed provide scores of such objects worthy of further investigation.
\end{itemize}

In summary, of the 3199 UV sources cataloged with positions within 
R~$\leq0.55^{\rm o}$ of their respective field centers, 192, or $6\%$, have no 
SDSS counterpart within $6\asec$. If we make some basic quality cuts on 
the {\it GALEX} data, this proportion does not change much: of the 2362 unflagged 
{\it GALEX} sources within the $0.55^{\rm o}$ radius, 130, or $5.5\%$, are not matched 
with an SDSS source. Finally, if we require that the sources have 
$n < 21.5$, there are 1537 {\it GALEX} sources within the $0.55^{\rm o}$ radius, and
only 21, or $1.4\%$, without an SDSS counterpart.

We can discount 10 of these 21 sources as probably artifacts based on their 
extraction flags. That leaves 11 UV sources out of 1537, or $0.7\%$, as 
photometric {\it GALEX} sources without an SDSS counterpart within $6\asec$. We 
have examined the {\it GALEX} and SDSS images for all 21 of the sources without a 
counterpart; in a handful of cases, we are unable to identify even an unlikely 
source (i.e., a distant star's halo) as responsible for the {\it GALEX} detection. 
While these comprise fewer than $0.3\%$ of the photometric {\it GALEX} 
sources, and are likely to be artifacts, they may be objects detected only in 
the UV and therefore of great interest.

\subsection{The Repeatability of {\it GALEX} Measurements}
The three {\it GALEX} AIS ERO fields overlap slightly. We therefore matched the {\it GALEX}
catalogs for the AIS fields with each other in order to characterize the 
differences between the measurements of objects observed twice. There are 31 
multiply observed {\it GALEX} sources that pass the quality cuts discussed above.  

The systematic astrometric offsets in both coordinates are consistent 
with {\it GALEX} astrometric errors inferred from comparison with SDSS astrometry. 
The root--mean--square (rms) scatter is somewhat larger ($2\asec$), probably 
because the multiply observed {\it GALEX} objects are detected near the edges of the 
fields.

The rms scatter for the $n$ band measurements is 0.33 magnitudes (only a small 
fraction of sources is detected in both bands both times). 
The magnitude differences depend on the mean $n$ magnitude for $n>20$.
For sources at the bright end (eight sources with $n<20$) we find that 
the median offset is 0.13 magnitudes, with an rms of only 0.07 magnitudes.
The magnitude difference normalized by the expected error has an  
rms scatter of 1.4, and 1.9 at the bright end. This
demonstrates that the photometric errors are computed fairly accurately
by the {\it GALEX} photometric pipeline, and that systematic errors at
the bright end are not very large.

\section{                    Analysis                       }

In this section we first compare the optical properties of matched
sources to the full SDSS sample, and then extend our analysis by
combining UV, optical, and IR data from the {\it GALEX}, SDSS, and 2MASS surveys.
The sample of matched sources analyzed here is UV--selected, since 
practically every {\it GALEX} source is detected by SDSS, while only 2.5\%
of SDSS sources are detected by {\it GALEX}. Not all {\it GALEX}/SDSS sources are 
detected by 2MASS (this is especially true for resolved sources; see 
Ivezi\'{c} et al.~2001), but this has no impact on the UV--optical--infrared 
SEDs discussed in \S 4.3.

SDSS color--magnitude and color--color diagrams are a powerful tool
to classify detected sources (e.g., Fan~1999, Finlator et al.~2000, 
Richards et al.~2002, and references therein), thanks to accurate five band 
photometry and robust star/galaxy separation. Thus, when studying a subsample 
of sources selected by other means, such as detections at non--optical 
wavelengths, it is very informative to examine their distribution in 
these diagrams.  

\begin{figure} 
\includegraphics[bb=0 127 542 739, width=0.85\columnwidth]{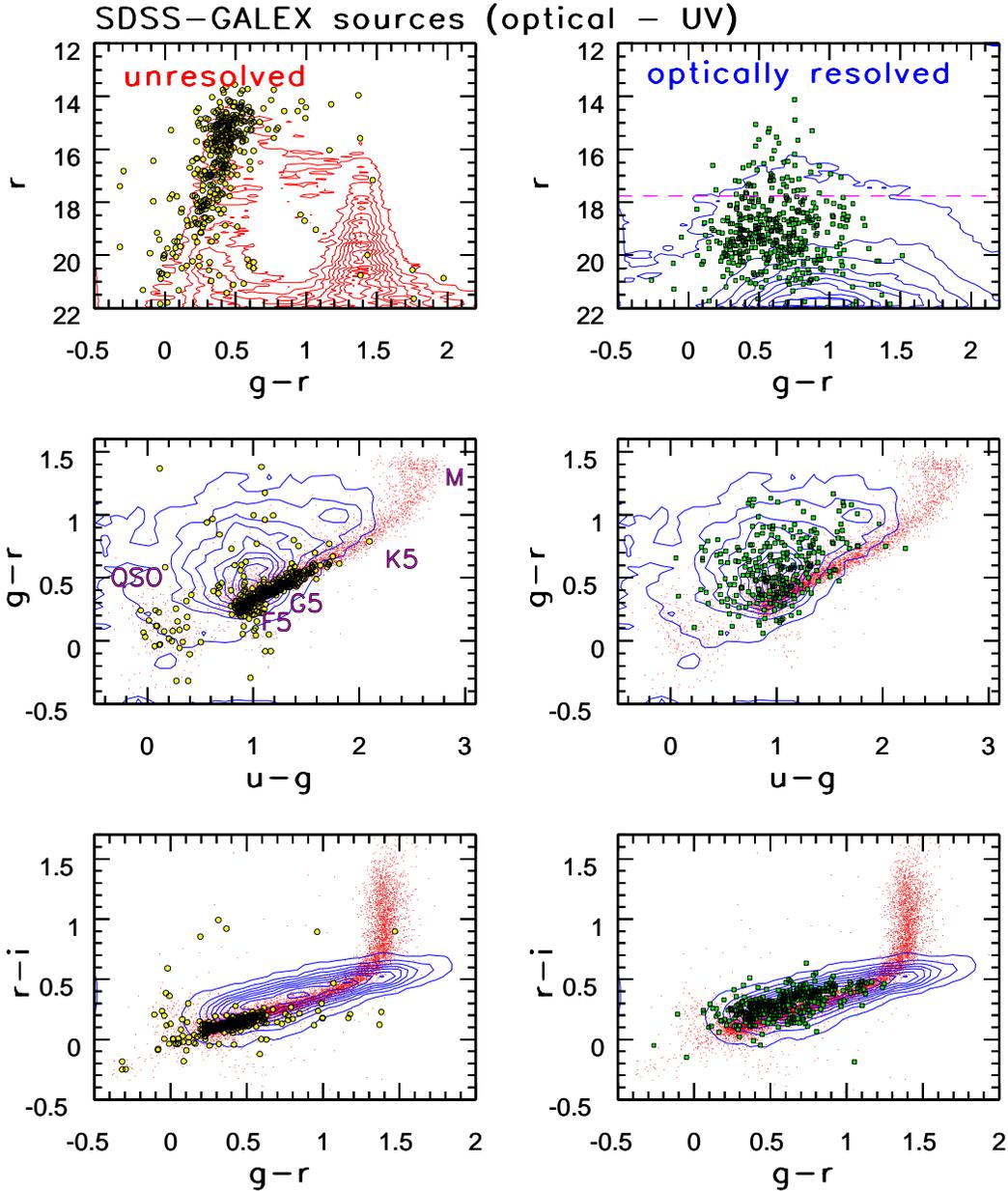}
\caption{The top two panels compare the distribution of SDSS sources 
detected by {\it GALEX} (symbols) to the overall distribution of all SDSS sources 
in the $r$ vs $g-r$ color--magnitude diagrams, for one of the three {\it GALEX}
AIS ERO fields discussed here. The left column corresponds 
to optically unresolved sources, and the right column to optically resolved 
sources. The bottom four panels show color--color diagrams, where the
distributions of all SDSS unresolved sources are shown by small dots, and 
those for galaxies by contours (same for the left and right panels). 
The {\it GALEX}/SDSS sources are marked by large symbols (unresolved left and 
resolved right). The approximate positions of low--redshift quasars and a few 
characteristic stellar spectral types are shown in the middle left panel.
The dashed line in the top right panel marks the faint flux limit
for the SDSS spectroscopic ``main'' sample ($r_{Pet}<17.8$).
\label{SDSScmd}}
\end{figure}

The contours in the top two panels of Fig.~\ref{SDSScmd} outline 
the distribution of optically unresolved (left) and resolved (right) 
SDSS sources in the $r$ vs $g-r$ color--magnitude diagram (we use the SDSS 
{\it model} magnitudes; for details see Stoughton et al.~2002). The matched
{\it GALEX}/SDSS sources are shown by symbols. For {\it GALEX} detections we require
$n<21$ or $f<21$ and correct magnitudes for interstellar extinction
using $A_f=2.97A_r$ and $A_n=3.23A_r$, where $A_r$ is the $r$ band
extinction from the maps of Schlegel, Finkbeiner, \& Davis (1998), 
distributed with the SDSS data. These coefficients were evaluated using
the standard interstellar extinction law\footnote{The standard Milky Way 
extinction curve predicts that the $f-n$ color becomes {\it bluer} with 
increasing extinction---this is a consequence of the strong feature at 
0.22 $\mu$m (e.g., Fig.~$21$ in Calzetti, Kinney, \& Storchi--Bergmann 1994).}
from Cardelli, Clayton, \& Mathis 
(1989; M.~Seibert 2004, private communication). The median $A_r$ for the 
three AIS fields is 0.12, with a root--mean--square scatter of 0.02 magnitudes.

The remaining panels in Fig.~\ref{SDSScmd} show the distribution of optically 
unresolved (dots) and resolved (contours) SDSS sources in the $g-r$ vs 
$u-g$ (middle row) and $r-i$ vs $g-r$ (bottom row) color--color diagrams, which
we discuss in the next two sections.

\subsection{   Unresolved SDSS Sources  }

The optically unresolved {\it GALEX}/SDSS sources are dominated by blue turn--off 
stars ($0.8 < u-g<1.5$ and $0.2 < g-r < 0.6$, see the middle left panel in
Fig.~\ref{SDSScmd}). The sample also contains 
low--redshift quasars ($z<2.2$) and hot white dwarfs (both are identified by their 
blue $u-g$ colors, $u-g < 0.6$), as well as white dwarf--M dwarf pairs (scattered 
above the locus;
for details, see Smol\v{c}i\'{c} et al.~2004 and Pourbaix et al.~2004). 
The well--defined red edge of the turn--off star distribution 
in the $r$ vs $g-r$ color--magnitude diagram (at $g-r\sim0.6$ for $r\sim14$ and
$g-r\sim0.2$ for $r\sim19$) is a consequence of the 
{\it GALEX} faint limit and the steep dependence of the UV--optical color on 
the effective temperature (the latter essentially controls the $g-r$ color).
For these stars we find that $n-r = f(g-r)=12.3\,(g-r)-0.47$, and thus the 
faint limit in the {\it GALEX} $n$ band ($n<21$) defines the observed red edge:
$r < 21.47 - 12.3\,(g-r)$.

The top left panel in Fig.~\ref{CCDs} shows the distribution of optically 
unresolved {\it GALEX}/SDSS sources in the $n$ vs $n-u$ color--magnitude diagram. 
Sources detected only in the {\it GALEX} NUV band are shown as small dots, and 
those with detections in both FUV and NUV bands as large dots. The easily 
discernible bimodal distribution of the $n-u$ color is well correlated with 
the distribution of the SDSS $u-g$ color, as shown in the top right panel. 
The boundary $n-u=1.3$ corresponds to $u-g=0.6$ which separates turn--off 
stars from hotter stars (T$_{eff}>10000$~K) and low--redshift quasars. 
The last two classes dominate the optically unresolved sources detected in 
both {\it GALEX} bands. 
As discernible from the middle left panel, the fraction of {\it GALEX}/SDSS sources 
detected in both {\it GALEX} bands is much higher for hot stars ($u-g < 0.6$, 
$g-r < -0.2$, dominated by white dwarfs) than for quasars 
($u-g < 0.6$, $g-r > -0.2$). This is a consequence of the {\it GALEX} faint limit
in the FUV band and the fact that the $f-n$ color is {\it bluer} for hot 
stars than for quasars (see the middle right panel in Fig.~\ref{CCDs}). 
In addition, quasars at redshifts beyond $\sim$$0.5$
may be very faint in the $f$ band because the Ly~$\alpha$ line is redshifted 
to the $n$ band. 

\subsection{           Resolved SDSS Sources        } 

The optically resolved {\it GALEX}/SDSS sources are dominated by galaxies 
{\it fainter} than the SDSS spectroscopic limit for the ``main" sample 
($r_{Pet}=17.8$\footnote{The SDSS Petrosian magnitude, $r_{Pet}$, is computed using 
the Petrosian flux. The Petrosian flux is measured in a circular aperture 
of radius twice the Petrosian radius, where the latter is defined by
the ratio of the averaged and local surface brightness. See Strauss et al.
(2002) for details.}), 
but mostly brighter than $r=21$, as 
discernible from the top right panel in Fig.~\ref{SDSScmd}. {\it GALEX}/SDSS 
galaxies are predominantly blue ($0.2 < g-r < 0.8$, or $u-r<2.2$; for a 
discussion of the bimodal $u-r$ color distribution of galaxies see 
Strateva et al.~2001), while a small fraction have colors 
consistent with those of AGN ($2 < u-r < 3$; Obri\'{c} et al.~2005, in preparation).

\begin{figure} 
\includegraphics[bb=0 127 542 739, width=0.85\columnwidth]{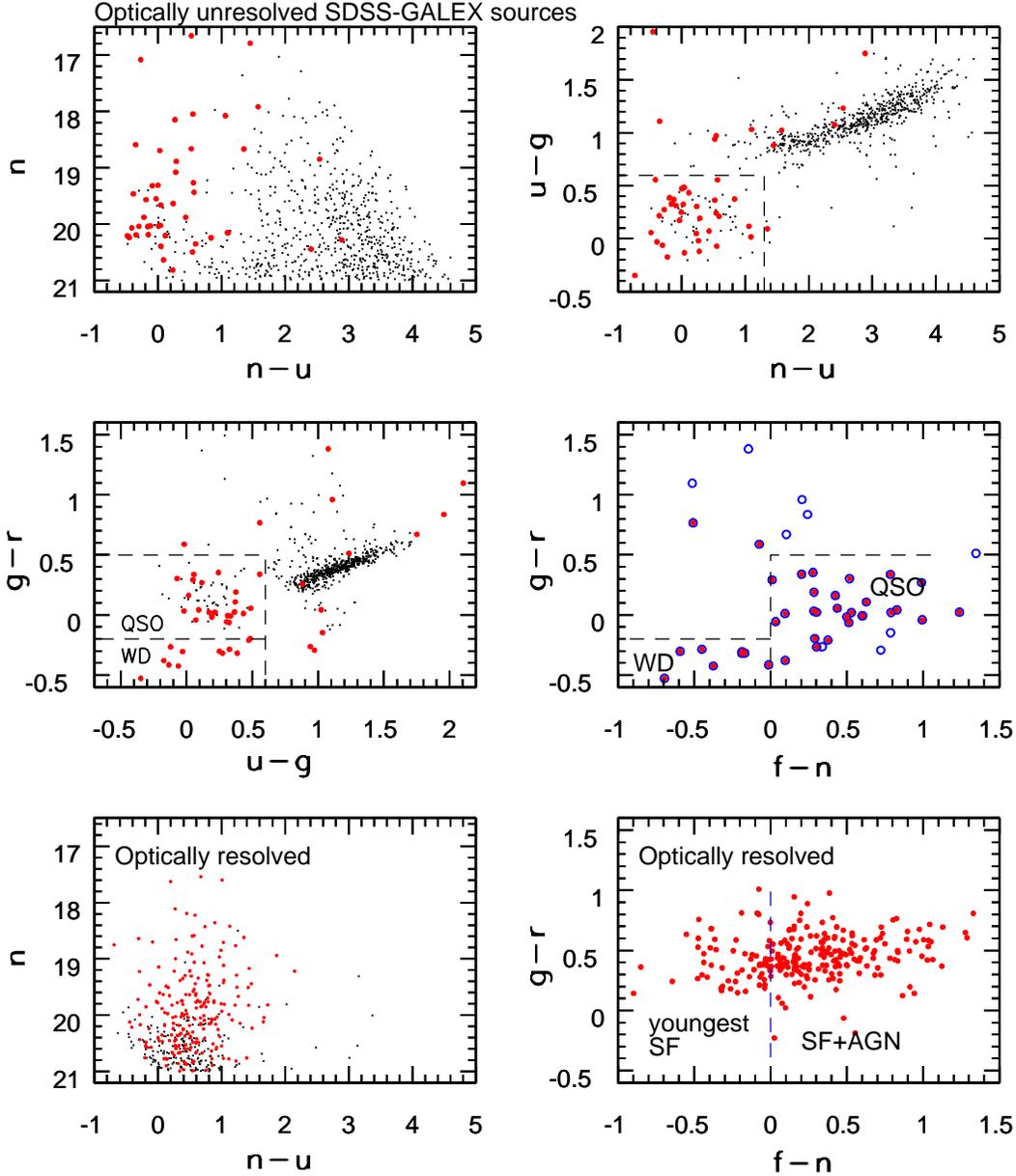}
\caption{{\it GALEX}/SDSS UV--optical color--color and color--magnitude diagrams
for all sources from the three {\it GALEX} AIS ERO fields discussed here.
The top four panels correspond to optically unresolved sources, and the
bottom two panels to galaxies. The small dots are sources detected only
in the {\it GALEX} $n$ band, and the large dots are those detected in both {\it GALEX}
bands. Circles in the middle right panel mark objects with $u-g>0.6$,
and large dots those with $u-g<0.6$ (among the latter, white dwarfs 
dominate for $g-r<-0.2$, and low--redshift quasars for $g-r>-0.2$); note 
that white dwarfs (WD) have bluer $f-n$ colors than quasars (QSO). The dashed
line in the bottom right panel separates galaxies with the youngest starbursts
(left) from those consistent both with intermediate age starbursts and with AGN 
emission (right), as inferred from comparison with the middle right panel.
\label{CCDs}} 
\end{figure}

The distribution of optically resolved {\it GALEX}/SDSS sources in the $n$ vs $n-u$ 
color--magnitude diagram is shown in the bottom left panel in Fig.~\ref{CCDs}, 
where those detected only in the {\it GALEX} NUV band are shown as small dots, and 
those with detections in both FUV and NUV bands as large dots. Unlike optically 
unresolved sources, whose detection in both {\it GALEX} bands is strongly correlated 
with the $n-u$ color, for optically resolved sources the fraction of those with 
detection in both {\it GALEX} bands is strongly correlated with brightness: galaxies 
brighter than $n=20.5$ typically have both detections, and those with only one 
detection are dominated by galaxies with $n>20.5$. The fairly narrow $n-u$ color
distribution suggests that the mismatching of SDSS and {\it GALEX} detections
and other problems such as the shredding of extended galaxies by the 
{\it GALEX} photometric pipeline discussed by Seibert et al.~(2005a) are not 
significant for this sample.

Galaxies having undergone recent starbursts have UV fluxes dominated by 
their most massive young stars. These hot stars have T$_{eff}>10000$~K 
and UV spectral slopes ($f-n$ colors) similar to those of hot white dwarfs. 
As these galaxies age, stellar evolution will preferentially remove the 
hottest, bluest members first, and their UV color will grow redder\footnote{According
to models by Bianchi et al.~(2005), the $f-n$ color changes from $-0.35$ to
$-0.06$ to $0.18$ as a single stellar population ages from 1 Myr to 10 Myr
and to 100 Myr.}. By comparison, the UV flux of AGN host galaxies is dominated 
by emission from their central source, whose UV spectral slope is similar to 
that of (unresolved) low--redshift quasars.

In the middle right panel of Fig.~\ref{CCDs}, we divide the $g-r$ vs $f-n$ 
color--color diagram for unresolved sources into regions dominated by hot 
white dwarfs ($f-n < 0$, $g-r < -0.2$) and by low--redshift quasars 
($f-n > 0$, $g-r > -0.2$). {\it GALEX} photometric errors should make negligible 
contributions to the observed color dispersion, as our flux limits are 
conservative ($n<21$ and $f<21$); nevertheless, some of the
extreme color outliers could reflect non--Gaussian errors, such as the
pipeline's treatment of complex or blended sources, or {\it GALEX}/SDSS mismatches.
In addition, dust attenuation may affect the integrated $f-n$ color of galaxies,
and bias the implied stellar ages discussed below towards larger values.
Using the model results from Salim et al.\ (2005), we estimate that the 
median reddening of the $f-n$ color due to dust may be about 0.5 magnitudes. For 
this reason, we emphasize that the adopted $f-n=0$ boundary is intrinsically fuzzy.

As shown in the bottom right panel of 
Fig.~\ref{CCDs}, comparing the colors of {\it GALEX}/SDSS galaxies to those of the 
unresolved sources suggests that {\it GALEX}/SDSS galaxies with $f-n < 0$ are likely 
to be the youngest starburst galaxies, with UV colors still dominated
by flux from very hot stars (plausible ages, inferred from models, are 
less than $\sim$20 Myr; e.g., Bianchi et al.~2005). Furthermore, this sample 
should not suffer seriously from AGN contamination, as relatively few low--redshift 
quasars have $f-n$ colors this blue.

Resolved sources with $f-n > 0$, however, while consistent with a population 
of older starburst galaxies, may also contain a significant fraction of AGN hosts, 
given that low--redshift quasars share this UV color space. While {\it GALEX} $f-n$ colors 
will provide constraints on the star formation history with greater 
precision than is possible from SDSS data, since the {\it GALEX} $f-n$ color varies 
substantially more than the SDSS $g-r$ color ($\Delta(f-n)/\Delta(g-r)\sim4$), 
additional information, such as spatially resolved far--UV emission, or X--ray and 
radio data, is needed before the {\it GALEX} UV color can be unambiguously 
interpreted as a sign of recent star formation.
An analogous conclusion follows from the distribution of the $n-u$ color: for the 
majority of {\it GALEX}/SDSS galaxies $n-u$ is bluer than the $n-u$ color for turn--off 
stars in the Galaxy, and is similar to $n-u$ colors of {\it both} quasars 
and hot stars (compare the top left and bottom left panels in Fig.~\ref{CCDs}). 

\begin{figure} 
\centering
\includegraphics[bb=2 5 645 480, clip=,width=0.9\columnwidth]{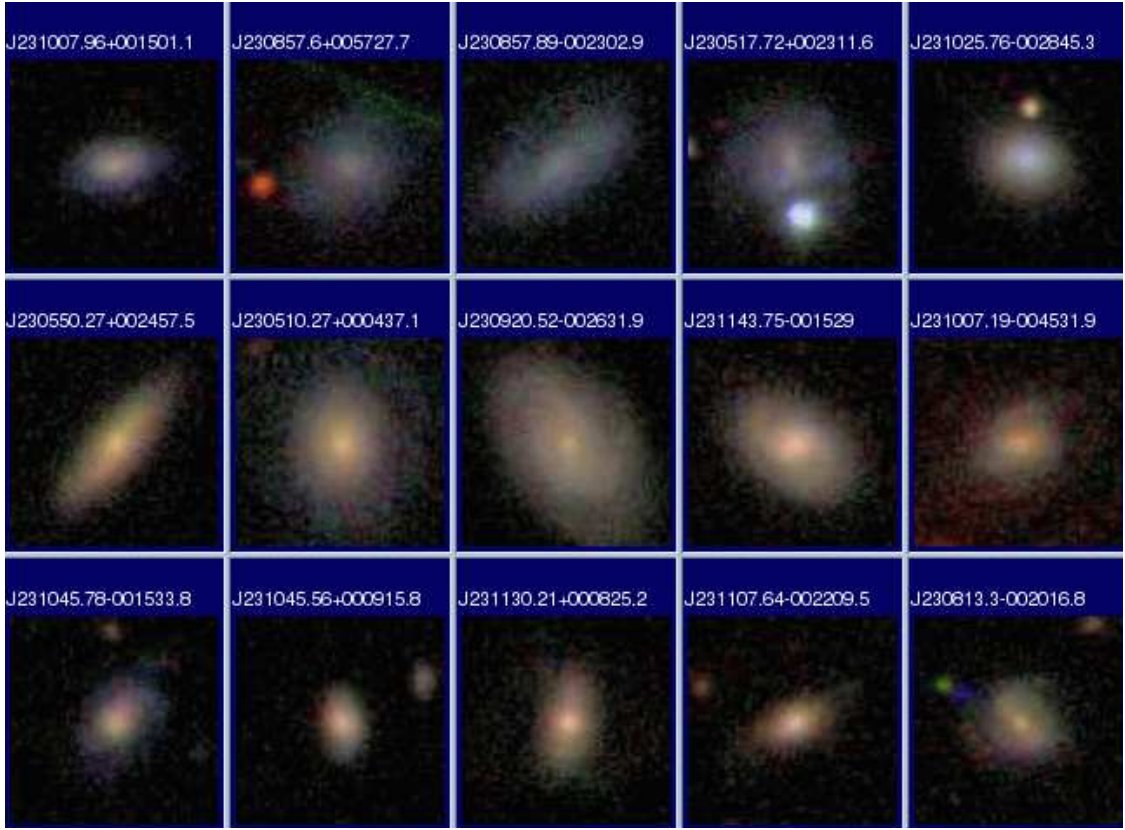}
\caption{{\it g, r, i} composite SDSS images of a randomly chosen sample of ``main'' galaxies detected by {\it GALEX} from among those discussed and classified by Obri\'{c} et al.~(2005, in preparation). The first row shows images of star--forming galaxies, the second of AGN; galaxies in the third row have uncertain classifications based on their emission line ratios. North is up, and the images are roughly $25\asec$ on a side.
\label{gal}} 
\end{figure}

Further evidence that a blue UV color for {\it GALEX}/SDSS galaxies does not
necessarily imply starburst emission comes from a detailed analysis
of emission line strengths measured from SDSS spectra.
Obri\'{c} et al.~(2005) study the multi--wavelength properties of SDSS
``main'' spectroscopic galaxies and find that about 40\% of them are
detected by {\it GALEX}. Of those, 70\% are emission--line galaxies, which
they classify as AGN, star--forming, or ``inconclusive'' using line 
strength ratios. They find that {\it at least 10\% of SDSS ``main'' galaxies
detected by {\it GALEX} have emission lines indicating an AGN, with the true
fraction possibly as high as 30\%}. We have visually inspected SDSS 
$g,r, i$ color composite images of these galaxies (a total of 55) and found 
that the classification based on emission line strengths is well correlated 
with morphology. SDSS images of a random subsample of Obri\'{c} et 
al.~{\it GALEX}/SDSS AGN, star--forming, and ``inconclusive'' galaxies are 
presented in Fig.~\ref{gal}, and show clear morphological differences between 
galaxies classified as star--forming and as AGN, with the latter being more 
centrally concentrated. These morphological differences further demonstrate 
that at least some {\it GALEX}/SDSS galaxies are more likely to be AGN than 
star--forming. In Table 3, we list Obri\'{c} et al.'s measurements of the 
light concentration indices (see Strateva et al.~2001 for details) and 
emission line strengths, SDSS redshifts, and {\it GALEX}, SDSS, and 2MASS 
photometry/colors, for the AGN candidates in Fig.~\ref{gal}. We note that
one of the AGN candidates, SDSS J230920.52$-$002631.9, is cataloged by SIMBAD 
as the Seyfert 1 galaxy [VV2003c] J230920.5$-$002632, while another, 
SDSS J231143.75$-$001529 is $<1\asec$ from a cataloged FIRST source 
(Becker, White, \& Helfand 1995). 

Finally, we note that although young stellar populations dominate the UV flux
from starburst galaxies, their contribution to the UV--optical--infrared flux is very 
small, as inferred from the red $g-r$ colors for these sources ($g-r \sim 0.3$, 
unlike $g-r \sim -0.4$ typical for stars with $f-n<0$). We discuss the 
contribution of UV light to the UV--optical--IR flux further below.

\subsection{     The 10 band UV--Optical--IR Spectral Energy Distributions }

In addition to color--color and color--magnitude diagrams, an efficient
way to analyze data that span such a wide wavelength range is to construct 
the spectral energy distributions (SEDs) for various classes of sources. Here 
we analyze the turn--off stars, hot stars, low--redshift quasars, and
two subsamples of galaxies. We expand the wavelength range by including 
2MASS data; we use the Vega--to--AB conversion for 2MASS magnitudes as 
described by Finlator et al.~2000: $J_{AB}=J_{2MASS}+ 0.89$, 
$H_{AB}=H_{2MASS}+ 1.37$, $K_{AB}=K_{2MASS}+ 1.84$. The SEDs are presented in the 
$\lambda\,F_\lambda$ (=$\nu\,F_\nu$) form, normalized to 1 at 2.2 $\mu$m 
(2MASS $K_S$ band).  

\begin{figure}
\includegraphics[bb=60 87 542 709, width=0.9\columnwidth]{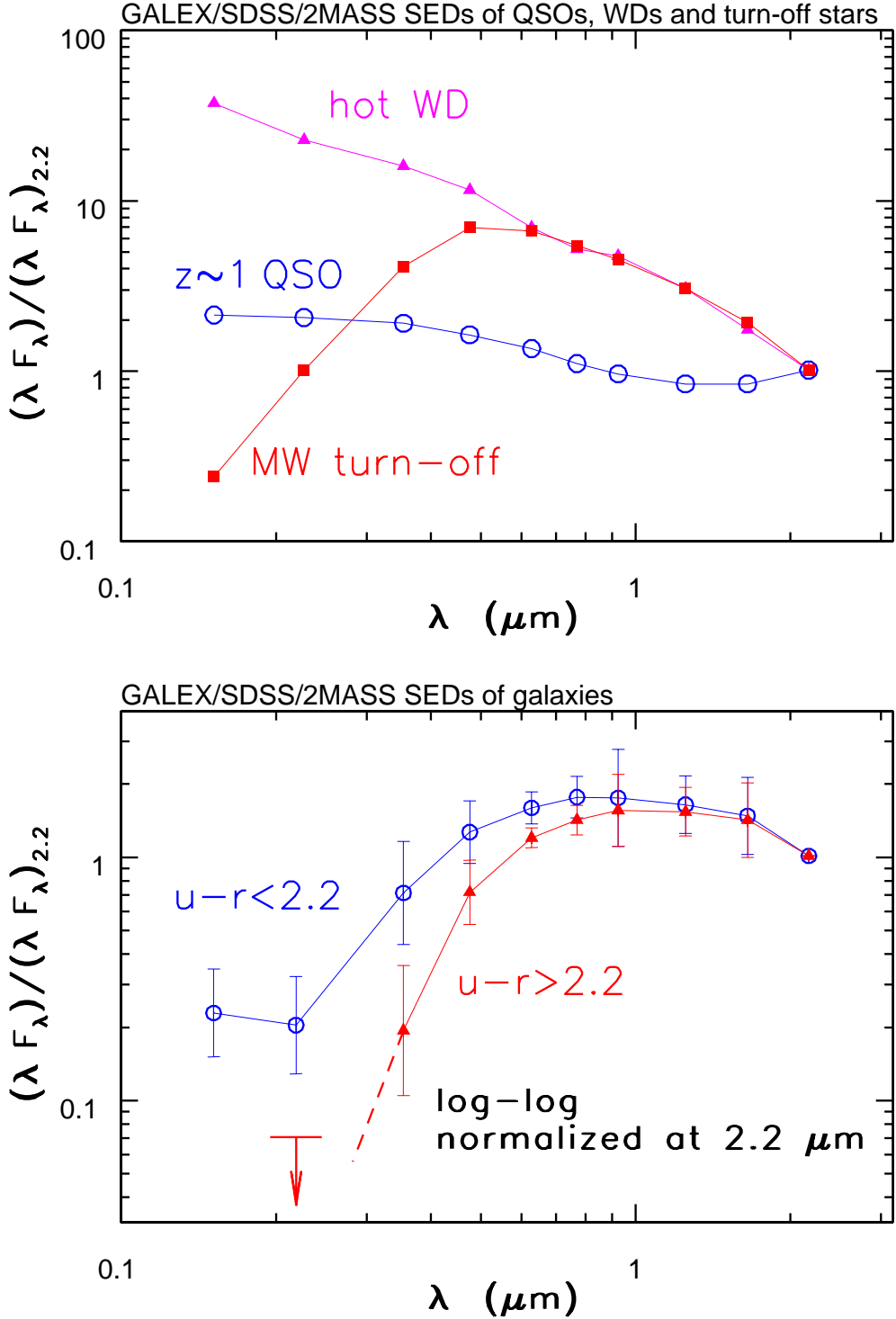}
\caption{The top panel shows median UV--IR SEDs for low--redshift ($z\sim1$) 
quasars (circles), hot white dwarfs (triangles) and turn--off stars (squares), 
constructed using {\it GALEX}, SDSS and 2MASS data (the data points are connected 
to guide the eye). The bottom panel shows the mean SEDs for
blue ($u-r<2.22$, circles) and red ($u-r>2.22$, triangles) galaxies, with 
redshifts in the range $0.03-0.05$. The error bars show the root--mean--square
color scatter for each subsample, determined from the interquartile range.}
\label{sedBR}
\end{figure}

For hot and turn--off stars, we select subsamples in the SDSS $g-r$ vs $u-g$ 
color--color diagram (see \S 4.1 and Fig.~\ref{SDSScmd}), and use the 
median colors (e.g., for $f-n$, $n-u$, \dots, $H-K_S$) 
to construct their SEDs. Optical colors of low--redshift quasars vary by
a few tenths of a magnitude as a function of redshift, due to emission line
effects (Richards et al.~2001). We adopt optical colors representative of 
objects at $z=1$ (i.e., roughly the median redshift). The sample of {\it GALEX}/SDSS 
low--redshift quasars discussed here is not sufficiently large to constrain the 
dependence of UV colors on redshift, and we simply adopt the median values for 
$f-n$ and $n-u$ colors. For 2MASS colors (which vary less as a function of 
redshift than do the optical colors), we take the median values of $z-J$, $J-H$ 
and $H-K_S$ colors for a sample of low--redshift quasars discussed by Covey et 
al.~(2005, in preparation; these values agree well with the results of Finlator et 
al.~2000). The SEDs for these three representative classes of optically 
unresolved sources are shown in the top panel in Fig.~\ref{sedBR}. Note that 
the well--known 1 $\mu$m inflection in the quasar SED (e.g., Elvis et al.~1994)
is properly reproduced. 

The observed broad--band colors of a galaxy depend both on its type and 
redshift (K correction). Due to the limited redshift range, the effect
of galaxy type dominates the observed color dispersion.
Following Strateva et al.~(2001), we separate galaxies in two dominant subsamples 
using the SDSS $u-r$ color; in practice, this roughly
corresponds to a morphological division into spiral and elliptical galaxies. 
The effect of K correction on measured optical and infrared galaxy colors 
is discussed in detail by Obri\'{c} et al.~(2005). Of the $99000$ ``main'' 
galaxies they study, 1880 blue and 3400 red galaxies listed 
in the 2MASS Extended Source Catalog and selected from the narrow redshift 
range $0.03<z<0.05$ are used to construct these SEDs. For the $u-r<2.2$ 
subsample we adopt the median $f-n$ and $n-u$ colors for the {\it GALEX}/SDSS galaxies
discussed here. For the $u-r>2.2$ subsample, only the $n-u$ median color is
used, while for the $f-n$ color we adopt a lower limit, based on the color
of the {\it GALEX} faint flux limits (most of those galaxies are not detected in the
$f$ band). The SEDs for the two dominant types of galaxies
are shown in the bottom panel in Fig.~\ref{sedBR}. The error bars
indicate the root--mean--square scatter in each color and for each subsample.

The comparison of the UV parts of SEDs for optically unresolved sources
and galaxies further illustrates the conclusions from the preceding section.
The very blue UV color for galaxies detected in both {\it GALEX} bands cannot
be due to stars with similar ages as the turn--off stars from the Galaxy. 
On the other hand, the observed UV slope is consistent with the UV slope
for both hot stars and low--redshift quasars. The contribution of the UV flux
to the UV--optical--infrared flux of galaxies is discussed next. 

\subsection{The UV Contribution to the UV--Optical--IR Flux of Galaxies}

Obri\'{c} et al.~(2005) present an analysis of the dependence of galaxy
SEDs on galaxy type. For each dominant galaxy type (defined by the $u-r$ color
division of Strateva et al.~2001) they compute the integrated flux in the 
$0.2-2.2$~$\mu$m range covered by {\it GALEX}, SDSS, and 2MASS data. Although we refer to 
this flux as the bolometric flux hereafter, note that it does not include the 
contributions from wavelengths longer than 2.2~$\mu$m, which, for galaxies with 
strong mid-- and far--infrared emission, could be as large, or larger, as those from 
the $0.2-2.2$~$\mu$m region (the contributions from wavelengths shorter 
than $0.2$~$\mu$m are most likely not important). 
Obri\'{c} et al.~demonstrate that galaxy SEDs, when normalized by this bolometric 
flux, cross at a wavelength corresponding to the SDSS $z$ band, regardless of 
the galaxy type. In other words, the bolometric correction for galaxies in 
the $z$ band is independent of type, and thus the $z$ band flux and absolute 
magnitude measurements are good proxies, to within a type--independent constant
(which they report as ($\lambda\,F_\lambda)_z = 0.58\,F_{bol}$), for bolometric 
flux and bolometric luminosity. Hence, the $f-z$ color
is a good choice for studying the UV contribution to the bolometric flux 
of galaxies. 

\begin{figure} 
\includegraphics[bb=0 127 542 709, width=0.85\columnwidth]{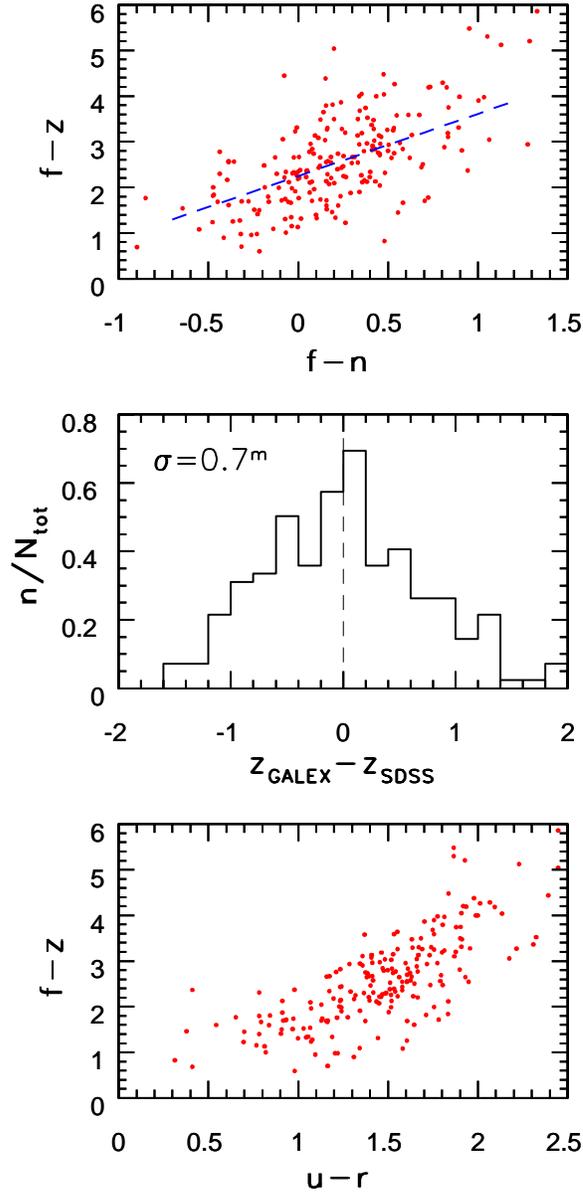}
\caption{The top panel shows the correlation between the $f-z$ (UV--IR color, 
a measure of the UV contribution to the UV--optical--infrared flux) and $f-n$ 
(UV slope) colors for galaxies detected in both {\it GALEX} bands. The dashed line is a 
best linear fit $f-z=1.36\,(f-n) + 2.25$ to the median $f-z$ color in bins of $f-n$. 
The distribution of differences between synthetic {\it GALEX}--based $z$ band 
magnitudes computed using this relation, and
the SDSS-measured $z$ band magnitudes is shown in the middle panel. The bottom
panel shows an apparently similar correlation between the $u-r$ and $f-z$
color. However, this correlation is a consequence of the strong selection effect
introduced by the {\it GALEX} faint flux limit in the $f$ band.  
\label{plot2}}
\end{figure}

The top panel in Fig.~\ref{plot2} shows the $f-z$ color of galaxies
detected in both {\it GALEX} bands as a function of the $f-n$ color. 
A good degree of correlation is evident: galaxies with bluest $f-n$ colors
also tend to have the bluest $f-z$ colors. The selection effects for the 
sample shown in Fig.~\ref{plot2} are simple and defined by the {\it GALEX} faint 
flux limits, since essentially all {\it GALEX} sources are detected by SDSS. 
Hence, the correlation between the $f-z$ and $f-n$ colors is an 
astrophysical relation, rather than, for example, a consequence of 
missing sources in the upper left and lower right corners due 
to faint flux limits (for a counterexample see below). In other words,
it is fair to use the {\it GALEX} $f$ and $n$ measurements to ``predict"
the SDSS $z$ magnitude. The relation $z=f-1.36(f-n)+2.25$,
shown by the dashed line in the top panel in Fig.~\ref{plot2},
predicts unbiased SDSS $z$ band magnitudes with a 
root--mean--square scatter of only 0.7 magnitudes (see the middle panel of 
Fig.~\ref{plot2}). 

This correlation probably includes both the effects of the age distribution of
stellar populations and dust attenuation effects. If the contribution 
of dust attenuation effects is not dominant\footnote{According to the effective extinction law from Calzetti, Kinney, \& Storchi--Bergmann (1994), $\Delta (f-z)/\Delta(f-n)=4.1$. Hence, even if that extinction law does not apply exactly  (e.g., if the dust is different, or if there are unaccounted for radiative transfer effects), the slope of the observed $f-z$ vs $f-n$ correlation (1.36) appears too small to be explained only by dust attenuation.}, then it implies that the hottest, and thus youngest, stellar populations 
seem to have a fair degree of knowledge about the older populations. A detailed 
study of this interesting possibility, including disentangling the contributions 
of stellar age and dust attenuation effects, and the transformation to more 
physical quantities like the current and integrated star--formation rate, is 
beyond the scope of this work and will be addressed elsewhere.

As an example of an apparent correlation between colors 
due to selection effects, we show the $f-z$ color as a function of the $u-r$ 
color in the bottom panel in Fig.~\ref{plot2}.  The distribution of galaxies
in this diagram represents a bivariate distribution of the UV
contribution to the UV--optical--IR flux (or luminosity) as a function
of the morphological type (i.e., the $u-r$ color). However, 
{\it only} those galaxies with substantial UV flux, relative
to the UV--optical--IR flux, are sufficiently bright to be detected by
{\it GALEX}. The sharp red $f-z$ cutoff in the distribution of 
sources, running from the lower left to the upper right corner, is therefore
a direct consequence of the {\it GALEX} faint flux limit, and does not represent
an intrinsic astrophysical correlation. This ``asymmetry"
with respect to the $f-z$ vs $f-n$ diagram discussed above comes 
from the fact that {\it every} {\it GALEX} galaxy is detected by SDSS, but 
only {\it some} SDSS galaxies (those with substantial star formation, 
or perhaps with AGN activity) are detected by {\it GALEX}. Equivalently,
the $f-z$ measurement is available for every galaxy with $f-n$ measurement,
but not for every galaxy with the $u-r$ measurement.

In the same way SDSS $z$ band magnitudes can be ``predicted'' from {\it GALEX} 
$f$ and $n$ measurements, the $f-z$ vs $u-r$ correlation can be used
to formally predict the $f$ band flux from the SDSS $u$, $r$ and $z$ 
band measurements, with a root--mean--square scatter of only 0.6 magnitudes. However,
this scatter is simply a measure of the slope of the differential $f$ 
magnitude distribution, just above the $f$ band faint cutoff. 
With several magnitudes deeper UV data, the apparent correlation 
in the bottom panel in Fig.~\ref{plot2} should disappear, and this
scatter would increase considerably.

\section{                       Discussion                }
This study, despite the relatively small sample of matched objects, indicates
the enormous potential of modern massive sensitive large--scale surveys, and
emphasizes the added value obtained by combining data from different 
wavelengths. The comparison of {\it GALEX} and SDSS data, as well as the analysis 
of repeated {\it GALEX} observations, demonstrates the high quality of the {\it GALEX} 
catalogs. We find no significant population of sources detected only by 
{\it GALEX}; the $\sim$1\% of {\it GALEX} sources without a probable SDSS counterpart 
appear to be dominated by processing artifacts. While the astrometric 
calibration seems to show systematic offsets of order $1\asec$, the reported 
photometric errors describe the behavior of {\it GALEX} photometry quite well.  

Although only 2.5\% of SDSS sources are detected by {\it GALEX}, the UV data carry 
important astrophysical information. For example, the {\it GALEX} measurements of 
the UV color allow much more accurate and robust estimates of star--formation 
history than possible using only SDSS data. However, we caution that the 
UV spectral slope for the majority of galaxies detected in both {\it GALEX} bands 
is consistent both with hot stars and with AGN activity. 
Additional information, such as spatially resolved far--UV emission, 
or X--ray, IR, and radio data, is needed before the blue {\it GALEX} UV color can be
unambiguously interpreted as a sign of recent star formation. For example, 
Yi et al.~(2005) interpreted the {\it GALEX} detections of 63 elliptical galaxies 
from an SDSS sample constructed by Bernardi et al.~(2003) as evidence for 
recent star formation. However, as their Fig.~3 shows, all 63 of those 
galaxies have $f-n>0$ (or $M_f-M_r>M_n-M_r$ in their nomenclature). Our work 
suggests that, at least in principle, their UV emission may instead reveal 
low--level AGN activity. 

Similarly to Yi et al., Rich et al.~(2005) analyze a sample of $\sim$$1000$
early--type {\it GALEX}/SDSS galaxies with redshifts $<$0.2. They select a subsample
of 172 quiescent early--type galaxies by excluding all those with any 
evidence for non early--type morphology, star formation, or AGN activity
(using emission lines), and point out a surprisingly large range of the 
$f-r$ color (from $\sim$3 to $\sim$8). We find that the observed range
of the $f-r$ color can be explained by a small contribution of AGN--like
emission to an otherwise normal (``old red and dead'') elliptical galaxy.
For example, assume that an AGN--like SED with $f-n=0$ and $n-r=0$ is
added to an elliptical galaxy SED with $f-n=2$ and $n-r=6$ (Gil de Paz et 
al.~2005), such that the AGN contribution to the $r$ band flux is $1\%$. 
The AGN contribution to the overall flux is then $70\%$ in the $n$ band
and $94\%$ in the $f$ band. That is, the $f-n$ color is dominated by
the AGN contribution and becomes 0.30 (with $f-r=4.9$). The addition
of such a low--level AGN emission would likely go undetected in SDSS spectra. 

We have used two special purpose analysis pipelines developed by Tremonti
et al.~(2004) and Hao et al.~(2005) to model and subtract the stellar
continuum and measure the residual emission lines in such composite 
AGN + galaxy spectra. Both codes produce
comparable results: the addition of an AGN--like SED with the continuum
contribution of $1\%$ in the $r$ band produces a signal--to--noise ratio
(SNR) for the H$\alpha$ emission line $>3$ in $25\%$ of galaxies,
and $>5$ in only $2\%$ of galaxies. When the SNR cutoff is imposed
on other lines needed to construct the BPT diagram (Baldwin, Phillips, \&
Terlevich 1981), such AGN emission is practically unnoticeable in SDSS
spectra, although it dominates the {\it GALEX} flux measurements! Hence, 
the UV emission from quiescent ellipticals discussed by Rich et al.\ could 
simply be due to low--level AGN activity. 

From our analysis of the UV colors of the low--redshift QSOs and the hottest 
stars detected by SDSS, we find that, in the absence of additional information,
the only robust criterion to avoid contamination by AGNs is to require $f-n<0$
(which, of course, biases the sample towards the youngest starbursts).
Indeed, Obri\'{c} et al.~(2005) use emission line strengths to separate 
star--forming from AGN galaxies in a sample of ``main'' SDSS spectroscopic 
galaxies detected by {\it GALEX}, and find that the median $f-n$ colors are 0.1 for 
star--forming galaxies and 0.5 for AGN galaxies, in good agreement with the 
analysis presented here. 

It should be noted that it is not obvious what exactly the $f-n$ color measures. 
For example, Seibert et al.~(2005b) tested the canonical UV color--attenuation 
(IRX--$\beta$) relation for starburst galaxies with a sample of {\it GALEX} and 
{\it Infrared Astronomical Satellite} ({\it IRAS}) galaxies, and found that 
it consistently overestimates the attenuation they derive from their sample by half a 
magnitude. While $f-n$ is certainly expected to be affected by dust attenuation 
(e.g., Kong et al.~2004, Buat et al.~2005), the distribution of galaxies in the 
$f-z$ vs $f-n$ diagram implies that the ages of the dominant stellar populations and 
the corresponding star--formation rates must play a significant role in determining 
the color of a galaxy (as opposed to simply reflecting a varying degree of reddening 
of one and the same intrinsic stellar population in different galaxies). 

Finally, models have some difficulties producing $f-n$ colors at the extreme 
blue edge $f-n < -0.5$ (e.g., Bianchi et al.\ 2005). While this discrepancy could 
of course point to suspect observations (e.g., non--Gaussian photometric errors), 
the modeling of far--UV colors of galaxies is notoriously difficult 
due to the unknown spatial distribution of dust and to the poorly constrained 
dust opacity in this wavelength range. Furthermore, the observed colors of 
hot stars in our Galaxy do extend all the way to $f-n <-0.5$.
 
In any case, we emphasize that most of our conclusions regarding the 
nature of {\it GALEX} sources are model--independent---for example, those that 
pertain to galaxies are based on the comparison of galaxy colors with those 
observed for Galactic sources and quasars using the same bands and the same 
instruments.   

\begin{acknowledgements}
{\noindent \bf Acknowledgments} We are grateful to Scott Anderson, Alberto 
Conti, Tim Heckman, Samir Salim, and Ted Wyder for their insights, to Mark Seibert for 
expert advice and for circulating a draft of his paper prior to publication,
and to Christy Tremonti for computing emission line strengths for
composite AGN + galaxy SEDs. We thank the anonymous referee for valuable
comments that helped to improve the paper.

Funding for the creation and distribution of the SDSS Archive has been provided by the Alfred P. Sloan Foundation, the Participating Institutions, the National Aeronautics and Space Administration, the National Science Foundation, the U.S. Department of Energy, the Japanese Monbukagakusho, and the Max Planck Society. The SDSS Web site is http://www.sdss.org/.

The SDSS is managed by the Astrophysical Research Consortium (ARC) for the Participating Institutions. The Participating Institutions are The University of Chicago, Fermilab, the Institute for Advanced Study, the Japan Participation Group, The Johns Hopkins University, the Korean Scientist Group, Los Alamos National Laboratory, the Max--Planck--Institute for Astronomy (MPIA), the Max--Planck--Institute for Astrophysics (MPA), New Mexico State University, University of Pittsburgh, Princeton University, the United States Naval Observatory, and the University of Washington.

This publication makes use of data products from the Two Micron All Sky Survey, which is a joint project of the University of Massachusetts and the Infrared Processing and Analysis Center/California Institute of Technology, funded by the National Aeronautics and Space Administration and the National Science Foundation.

The {\it Galaxy Evolution Explorer} ({\it GALEX}) is a NASA Small Explorer. The
mission was developed in cooperation with the Centre National d'Etudes
Spatiales of France and the Korean Ministry of Science and Technology.
\end{acknowledgements}

\clearpage

\clearpage

\clearpage

\clearpage

\clearpage

\clearpage

\clearpage

\clearpage

\clearpage

\clearpage

\clearpage

\clearpage
\begin{table}
\scriptsize
\begin{tabular}{lc|c|}
\cline{2-3} \multicolumn{1}{c|}{} & RA$_{{\it GALEX}}-$RA$_{\rm SDSS}$ & Dec$_{{\it GALEX}}$$-$Dec$_{\rm SDSS}$ \\
\multicolumn{1}{c|}{{\it GALEX} field center}& Median $\pm$$\sigma$ (arcsec) & Median $\pm$$\sigma$ (arcsec) \\
\hline
\hline
 & \multicolumn{2}{c}{{\it All matches}} \\
\hline
\multicolumn{1}{|l|}{$23^h06^m00'72$ $-00^{\rm o}06\amin32\asec4$} & $-0.88\pm1.95$ & $0.19\pm1.88$ \\
\multicolumn{1}{|l|}{$23^h07^m55'21$ $+00^{\rm o}39\amin57\asec6$} & $-0.44\pm1.83$ & $-0.10\pm1.89$ \\
\multicolumn{1}{|l|}{$23^h10^m14'64$ $-00^{\rm o}11\amin49\asec2$} & $-0.44\pm1.90$ & $-0.38\pm1.86$ \\
\hline
\multicolumn{1}{|r|}{{\it Combined fields}} & $-0.55\pm1.90$ & $-0.13\pm1.88$ \\ 
\hline
 & \multicolumn{2}{c}{{\it Matches with R $\leq0.55^{\rm o}$}} \\
\hline
\multicolumn{1}{|l|}{$23^h06^m00'72$ $-00^{\rm o}06\amin32\asec4$} & $-0.99\pm1.80$ & $0.27\pm1.74$ \\
\multicolumn{1}{|l|}{$23^h07^m55'21$ $+00^{\rm o}39\amin57\asec6$} & $-0.55\pm1.62$ & $-0.02\pm1.74$ \\
\multicolumn{1}{|l|}{$23^h10^m14'64$ $-00^{\rm o}11\amin49\asec2$} & $-0.55\pm1.76$ & $-0.32\pm1.64$ \\
\hline
\multicolumn{1}{|r|}{{\it Combined fields}} & $-0.66\pm1.73$ & $-0.07\pm1.68$ \\
\hline
 & \multicolumn{2}{c}{{\it Clean matches}} \\
\hline
\multicolumn{1}{|l|}{$23^h06^m00'72$ $-00^{\rm o}06\amin32\asec4$} & $-1.10\pm1.24$ & $0.24\pm1.20$ \\
\multicolumn{1}{|l|}{$23^h07^m55'21$ $+00^{\rm o}39\amin57\asec6$} & $-0.66\pm1.30$ & $0.00\pm1.18$ \\
\multicolumn{1}{|l|}{$23^h10^m14'64$ $-00^{\rm o}11\amin49\asec2$} & $-0.55\pm1.18$ & $-0.31\pm1.22$ \\
\hline
\multicolumn{1}{|r|}{{\it Combined fields}} & $-0.66\pm1.25$ & $-0.06\pm1.21$ \\ 
\hline
\end{tabular}
\caption{Positional offsets between {\it GALEX} and SDSS sources.}
\end{table}

\clearpage
\newpage
\begin{table}
\scriptsize
\begin{tabular}{|l|c|c|c|l|}
\hline
\multicolumn{1}{|c|}{{\it GALEX} name}& $n$ & $f$ & Flags? & \multicolumn{1}{c|}{Likely SDSS counterpart} \\ \hline
\hline
 \multicolumn{3}{c}{{\it Galaxies}} & \multicolumn{2}{c}{}\\ 
\hline
J230644.65$+$001302.13 & $21.43$ & N/A & NUV = 3 &  J230645.4$+$001309.5, $r= 15.94$, D $= 13.5 \asec$ \\
J230734.52$-$001731.04 & $20.04$ & $19.64$ & FUV, NUV = 3 &  J230734.4$-$001737.3, $r = 14.87$, D $= 6.5\asec$ \\
J230919.65$+$004515.64 & $18.20$ & $19.97$ & FUV, NUV = 3 &  J230920.2$+$004523.3, $r = 13.95$, D $= 11.2\asec$ \\
\hline
 \multicolumn{3}{c}{{\it Likely stellar artifacts}} & \multicolumn{2}{c}{}\\ 
\hline
J230518.70$-$002816.29 & $20.06$ & N/A & NUV = 1 & J230519.2$-$002741.3, $r = 10.86$, D $= 35.8\asec$ \\
J230717.62$-$001953.40 & $20.74$ & N/A & None & J230715.31$-$002008.8, $r = 14.01$, D $= 37.8\asec$ \\
J230751.11$+$003936.81 & $21.09$ & N/A & NUV = 2 & J230752.1$+$003858.5, $r = 13.32$, D $= 41.1\asec$ \\ 
J230852.36$-$001005.47 & $21.19$ & N/A & None & J230853.7$-$000942.1, $r = 9.85$, D $= 30.8\asec$ \\
J230959.96$-$003441.17 & $21.18$ & N/A & NUV = 1 & J231002.3$-$003433.1, $r = 11.20$, D $=36.0\asec$ \\ 
J231042.50$-$002126.92 & $21.20$ & N/A & None & J231041.6$-$002133.3, $r = 13.11$, D $= 14.9\asec$ \\
\hline
 \multicolumn{3}{c}{{\it Likely stellar artifacts?}} & \multicolumn{2}{c}{}\\ 
\hline
J230740.51$+$003458.86 & $21.23$ & N/A & None &  J230731.4$+$003538.1, $r = 11.68$, D $= 2.4\amin$  \\
J230750.47$+$004018.21 & $21.22$ & N/A & None &  J230752.1$+$003858.5, $r = 13.32$, D $=1.4\amin$ \\
J230752.10$+$004008.03 & $21.31$ & N/A & None &  J230752.1$+$003858.5, $r = 13.32$, D $= 1.2\amin$ \\
J230754.36$+$000907.62 & $21.00$ & N/A & NUV = 2 &  J230800.1$+$000710, $r = 12.35$, D $=2.4\amin$ \\
J230756.56$+$000859.81 & $20.58$ & N/A & NUV = 2 &  J230800.1$+$000710, $r = 12.35$, D $= 2.0\amin$ \\ 
J230911.58$+$002631.83 & $20.74$ & N/A & None &  J230923.4$+$002727.5, $r = 9.56$, D $= 3.1\amin$ \\
J230943.69$+$000822.11 & $20.94$ & N/A & None & J230946.4$+$001041.5, $r = 10.95$, D $=2.4\amin$ \\
J231015.82$-$004246.95 & $21.46$ & N/A & NUV = 2 & J231017.9$-$004123.5, $r = 12.14$, D $=1.5\amin$ \\
\hline
 \multicolumn{3}{c}{{\it Unexplained}} & \multicolumn{2}{c}{}\\ 
\hline
J231000.10$-$001636.82 & $20.19$ & N/A & None & \\ 
J231011.21$-$001211.11 & $20.90$ & N/A & None & \\ 
J231011.36$-$001227.72 & $20.76$ & N/A & None & \\
J231131.21$-$002510.96 & $20.77$ & N/A & NUV = 2 & \\
\hline
\end{tabular}
\caption{{\it GALEX} objects without an SDSS counterpart within 6$\asec$.}
\end{table}

\clearpage
\newpage
\begin{table}
\scriptsize
\begin{tabular}{|l||c|c|c|c|c|}
\hline
SDSS name & J230550.27 & J230510.27 & J230920.52 & J231143.75 & J231007.19 \\
 & $+$002457.5 & $+$000437.1 & $-$002631.9 & $-$001529 & $-$004531.9 \\
\hline
\hline
RA & $346.459$ & $346.293$ & $347.336$ & $347.932$ & $347.530$ \\
Dec & $0.416$ & $0.077$ & $-0.442$ & $-0.258$ & $-0.759$ \\
CI & $2.40$ & $2.68$ & $2.03$ & $2.40$ & $2.18$ \\
$[$NII$]$/H$_\alpha$ & $-0.19$ & $-0.32$ & $-0.13$ & $-0.16$ & $0.06$ \\
$[$OIII$]$/H$_\beta$ & $-0.08$ & $-0.11$ & $0.56$ & $0.48$ & $-0.23$ \\
z & $0.062$ & $0.056$ & $0.035$ & $0.060$ & $0.111$ \\
\hline
$f-n$ & $-0.08$ & $0.48$ & $0.54$ & $0.73$ & $1.28$\tablenotemark{a} \\
$n$ & $20.23$ & $18.36$ & $18.77$ & $18.71$ & $18.64$ \\
\hline
$u-g$ & $1.59$ & $1.47$ & $1.39$ & $1.28$ & $4.01$ \\
$g-r$ & $0.80$ & $0.67$ & $0.62$ & $0.64$ & $0.81$ \\
$r-i$ & $0.42$ & $0.37$ & $0.32$ & $0.37$ & $0.34$ \\
$i-z$ & $0.30$ & $0.24$ & $0.23$ & $0.22$ & $0.37$ \\
$r$ & $16.37$ & $15.34$ & $15.53$ & $15.76$ & $17.53$ \\
$A_r$ & $0.12$ & $0.14$ & $0.10$ & $0.13$ & $0.10$ \\
\hline
$J-K_S$ & $1.10$ & $1.11$ & $1.05$ & $1.00$ & $1.06$ \\
$J$ & $15.53$ & $15.59$ & $16.00$ & $14.85$ & $16.13$ \\
$H$ & $14.88$ & $14.89$ & $14.11$\tablenotemark{b} & $14.21$ & $15.54$ \\
\hline
Comments & & LEDA 1156491 & Seyfert 1 & FIRST source & \\
\hline
\end{tabular}
\tablenotetext{a}{A nearby $r = 10$ star (saturated in SDSS) is likely to have affected the {\it GALEX} $n$ and SDSS $u$ measurements}
\tablenotetext{b}{Upper limit}
\caption{SDSS measurements of light concentration indices (CI), emission line strengths, and redshifts, along with {\it GALEX}/SDSS/2MASS photometry/colors, for the five {\it GALEX}/SDSS galaxies classified as AGN based on their emission line strengths and presented in Fig.~\ref{gal} (for details and references see Obri\'{c} et al.\ 2005).} 
\end{table}

\end{document}